\documentclass{INTERSPEECH2023}

\usepackage{textgreek}
\usepackage{upgreek}

\usepackage{multirow}
\usepackage[table,xcdraw]{xcolor}

\usepackage{float}

\interspeechcameraready

\title{Causal Signal-Based DCCRN with Overlapped-Frame Prediction\\ for Online Speech Enhancement}

 \name{Julitta Bartolewska, Stanisław Kacprzak, Konrad Kowalczyk%
}  

\address{AGH University of Science and Technology, Institute of Electronics, 30-059 Krakow, Poland}
\email{\{bartolew, skacprza, konrad.kowalczyk\}@agh.edu.pl}

\begin{document}

\maketitle
 
\begin{abstract}
The aim of speech enhancement is to improve speech signal quality and intelligibility from a noisy microphone signal. In many applications, it is crucial to enable processing with small computational complexity and minimal requirements regarding access to future signal samples (look-ahead). This paper presents signal-based causal DCCRN that improves online single-channel speech enhancement by reducing the required look-ahead and the number of network parameters. The proposed modifications include complex filtering of the signal, application of overlapped-frame prediction, causal convolutions and deconvolutions, and modification of the loss function. Results of performed experiments indicate that the proposed model with overlapped signal prediction and additional adjustments, achieves similar or better performance than the original DCCRN in terms of various speech enhancement metrics, while it reduces the latency and network parameter number by around $30\%$. %

\end{abstract}

\section{Introduction}
\label{sec:intro}

 Intelligibility and quality of the speech signal diminishes in presence of background noise. The aim of speech enhancement is to reduce this undesired effects and extract clean speech signal from a noisy mixture. Over the years, speech enhancement methods (both single- and multi-channel) based on deep learning turned out to be very effective for this task~\cite{nossier_supervised_se, wang_supervised_ss}, and nowadays are considered state-of-art. Those approaches can be broadly categorized into time and time-frequency domain methods. Time domain methods try to map noisy speech to clean speech directly \cite{Rethage2018wavenet, Pandey2019tcnn}%
 , while the ones operating in time-frequency domain usually define a learning target 
as clean speech spectogram or the desired mask (e.g., an ideal power ratio mask). Many
of them are based only on magnitude features (and re-use phase of a noisy signal)~\cite{heymann_2016, chakrabarty_2019}, however, methods that try to reconstruct phase using e.g. complex ratio mask have recently gained on popularity \cite{dcunet,dccrn,fullsubnet}.

Speech enhancement has many real-life applications, but the most desired ones require low processing latency, so that enhancement can be performed in real-time (without breaking the human communication process with unnatural delays). The low-computational cost is a property desired in all systems, but most crucial for on-device deployment on smartphones or hearing aids. These translate into \emph{algorithmic latency} and \emph{hardware latency} that both add up to the overall \emph{processing latency} \cite{wang2022stft,lowLatency2023survey}. The implementation aspects of speech enhancement are important parts of current research. Reduction of \emph{hardware latency} can be achieved by decreasing the number of network parameters (either by specific network (re-)design \cite{romaniuk20_interspeech}, by performing pruning \cite{tan2021compressing} or knowledge distillation \cite{thakker22_interspeech}) or their quantization \cite{lin2021seofp}. The main impact on \emph{algorithmic latency} has the length of speech frame that the system aims to predict, as single-frame prediction is the most popular approach for online speech enhancement. The frame is predicted based on current and previous inputs and overlap-added with previous predictions for final enhanced speech signal generation.
The recent challenges that focus on real-time speech enhancement, such as Interspeech 2020 Deep Noise Suppression (DNS) challenge \cite{dns2020} and Clarity \cite{clarity21}, put tight requirements on the maximum processing time and allowed look-ahead limited to respectively 40\,ms and 5\,ms. A comprehensive survey on recent advancements in low-latency speech enhancement can be found in \cite{lowLatency2023survey}.

In this work, we focus on a single architecture, namely Deep Complex Convolutional Recurrent Network (DCCRN) \cite{dccrn} (that has already shown impressive performance at DNS challenge), which originally estimates the so-called Complex Ratio Mask (CRM). In original DCCRN formulation, the estimated mask is applied to both real and imaginary parts, which can lead to improved signal enhancement due to phase information preservation. We analyze in-depth different modifications to the network architecture, which are aimed to improve speech enhancement performance in an online scenario. To this end we explore signal-based filtering instead of a mask-based approach, overlapped-frame prediction algorithm with partial and full sub-frames summation~\cite{Wang_2022}, causal architecture design (causal convolutions and deconvolutions), as well as loss modification and other subtle changes in the neural network architecture (described in detail in Sec. \ref{sec:method}).

The proposed causal signal-based filtering using DCCRN with full overlapped-frame prediction offers significant reductions in the number of neural network parameters,
as well as in the relative latency over the original mask-based DCCRN \cite{dccrn}. Furthermore, the results of experimental evaluations performed using Librimix and DNS datasets indicate similar yet often slightly better speech enhancement performance in terms of classical evaluation metrics.

\section{Proposed Causal DCCRN with Overlapped-Frame Prediction}
\label{sec:method}

\subsection{Problem formulation}
In this paper, we address the problem of single-channel speech enhancement performed in the short-time Fourier transform (STFT) domain. We assume an additive signal model in which the microphone mixture is given by  
\begin{equation}
Y^{(f,t)} = S^{(f,t)} + N^{(f,t)} \, ,
\end{equation}
where $Y^{(f,t)} = [\mathbf{Y}]_{f,t}$ denotes the $(f,t)$-th element of the matrix with a complex spectrum of the microphone signal $\mathbf{Y} \in \mathbb{C}^{F \times T}$, with the frequency and time indices given by $f=0,1,\ldots,F-1$ and $t=0,1,\ldots,T-1$, respectively. $S^{(f,t)} = [\mathbf{S}]_{f,t}$ and $N^{(f,t)} = [\mathbf{N}]_{f,t}$ are defined similarly and denote the source and noise signal spectra in the $(f,t)$-th time-frequency bin.

The goal of speech enhancement is to extract the desired source signal $\mathbf{s}$ in the time domain. In this work, it is obtained via inverse STFT of the source spectrum estimated by direct filtering of the complex STFT representation of the microphone signal by a complex-valued neural network, which can be written as
\begin{equation}
\mathbf{\hat{s}} = \mathcal{F}^{-1}\{\mathbf{\hat{S}} = H(\mathbf{Y}) \} \, ,  
\end{equation}
where $H(\cdot)$ represents neural network complex filtering and $\mathcal{F}^{-1}\{\cdot\}$ denotes an inverse STFT (ISTFT).
Note that the proposed direct filtering differs from the existing DCCRN-based approaches~\cite{dccrn} which estimate complex time-frequency masks.

\subsection{Direct signal filtering vs mask-based approach}
\label{signal_filtering}

\begin{figure}[!t]
\centerline{\includegraphics[width=0.85\columnwidth]{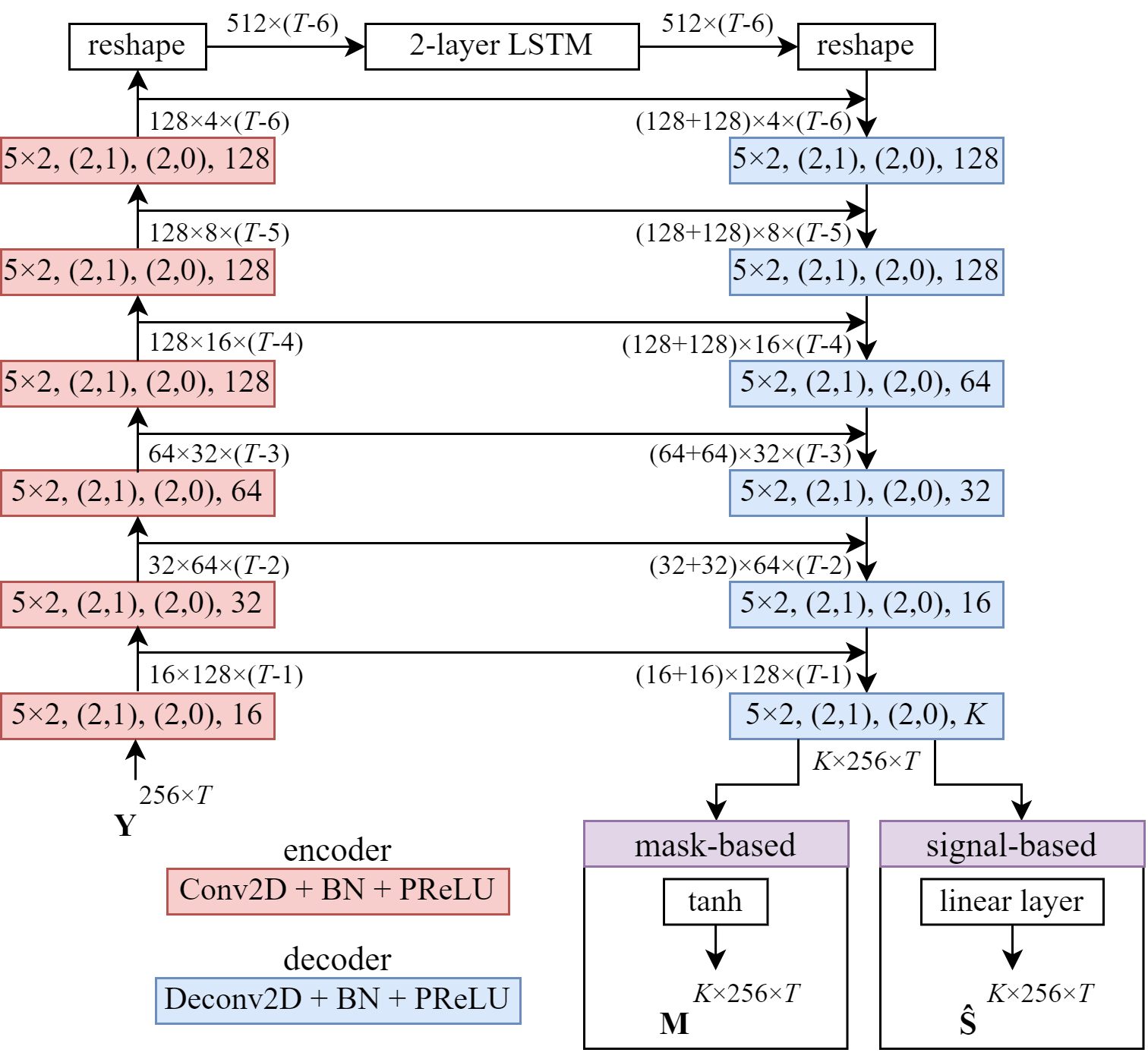}}
\caption{Original mask-based DCCRN architecture \cite{dccrn} and proposed signal-based, presented for overlapped-frame prediction. Tensor shapes are presented in the format: Features$\times$Freq$\times$Time, while Conv2D and Deconv2D blocks are shown as: kernelFreq$\times$kernelTime, (strideFreq, strideTime), (padFreq, padTime), features.}
\label{fig:org_dccrn}
\end{figure}

The original DCCRN \cite{dccrn} is a complex-valued network with encoder-decoder U-shaped architecture. Encoder and decoder are composed of six Conv2D/Deconv2D blocks, each consists of convolutional/deconvolutional layer followed by batch normalization and PReLU activation. It is designed with a 2-layer complex LSTM between encoder and decoder, with 128 hidden units, followed by a single linear layer with 512 units. Note that the provided layer dimensions (including those depicted in Fig. \ref{fig:org_dccrn}) are the same for the real and imaginary parts. Moreover, the Nyquist frequency is omitted in DNN-based processing.

State-of-the-art DCCRN~\cite{dccrn} estimates the so-called Complex Ratio Mask (CRM) $\mathbf{M} \in \mathbb{C}^{F \times T}$%
, which is next subject to complex multiplication with the complex spectrum of the microphone signal in the STFT domain.

In contrast, in this work, we directly perform complex filtering of the microphone signal spectrum by the DCCRN. As shown in Fig. \ref{fig:org_dccrn}, this can be achieved by minor modifications to the original 
network architecture. In particular, in comparison with the mask-based processing, the $\tanh(\cdot)$ activation function that limits the mask magnitude to the range $[0,1]$ is removed, and instead, a linear (i.e. fully connected) layer is added at the decoder output.

\subsection{Processing of overlapped time frames}
\label{overlapped}
For both mask-based and signal-based DCCRN, we apply the overlapped-frame prediction algorithm recently proposed in \cite{Wang_2022}, which allows to better leverage the future context without affecting their algorithmic latency. For each STFT frame, it assumes the prediction of $K=W/P$ frames, where $W$ and $P$ denote the frame length and the hop-size in samples, respectively. 
The predicted $K-1$ immediate past frames and the current one are then utilized within the ISTFT algorithm. After calculating, for each of $K$ predicted frames, the inverse transform, and applying the synthesis window, the resulting $K$ %
predictions for the frame in the time domain are finally appropriately overlap-added in order to produce the particular sub-frame of the output signal. This way, the so-called partial sub-frame summation is performed. In order to leverage all frame predictions, which involve a particular sub-frame, the so-called full sub-frame summation can be used. Note that the standard single-frame prediction is equivalent to the standard overlap-add method, which results as a special case of the full sub-frame summation for $K=1$. 

To ensure perfect reconstruction we employ the synthesis window design as presented in \cite{Wang_2022}.
In particular, for partial sub-frame summation, it is the same as the regular synthesis window used for single-frame prediction (since the summation process is mathematically equivalent), hence it follows
\begin{equation}
l[n] = \frac{g[n]}{\sum_{e=0}^{W/P-1}g[eP + (n \, \mathrm{mod} \, P)]^2} \, ,
\end{equation}
where $g$ is the analysis window and $0 \leq n < W$. However, for full sub-frame summation, it is modified to
\begin{equation}
l[n] = \frac{g[n]}{\sum_{e=0}^{W/P-1}\bigl((e+1) \times g[eP + (n \, \mathrm{mod} \, P)]^2\bigr)} \, .
\end{equation}

\begin{figure}[!t]
\centerline{\includegraphics[width=0.8\columnwidth]{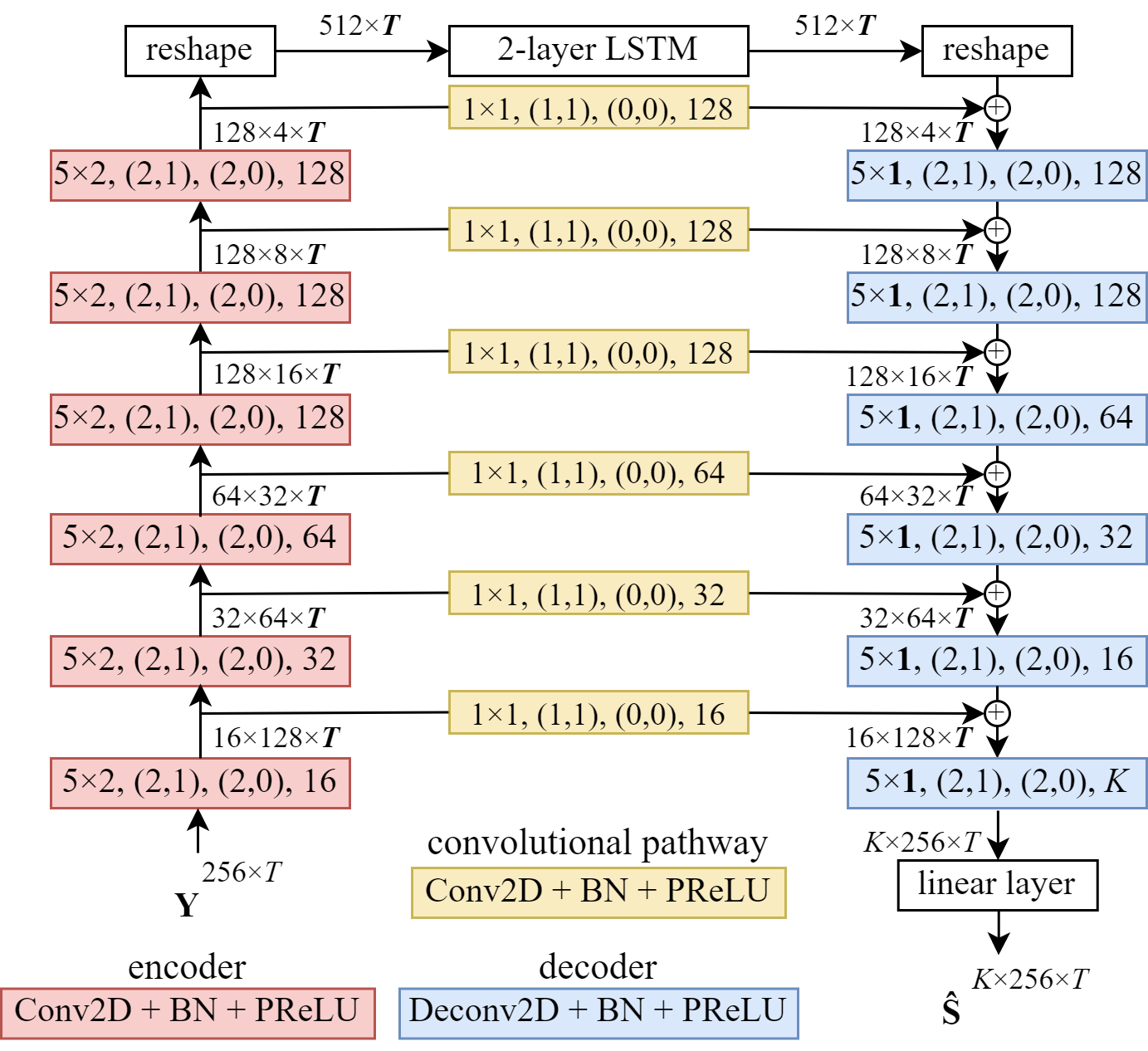}}
\caption{Proposed causal DCCRN with convolution pathways.}
\label{fig:causal_dccrn_cp}
\end{figure}

\subsection{Causal architecture design}
\label{casual}
In order to avoid reliance on future frames, %
non-causal layers are replaced with their respective causal counterparts. In particular, performed modifications include: (i) using causal two-dimensional (2D) convolutions in the encoder, achieved by the appropriate left zero-padding along the time dimension, %
and (ii) using causal two-dimensional deconvolutions in the decoder achieved by reducing the kernel size along the time dimension to the value of $1$. %

Note that as a result, in contrast to the original DCCRN~\cite{dccrn}, in both encoder and decoder, the size of the time dimension $T$ remains unchanged, as marked in Fig. \ref{fig:causal_dccrn_cp}.
Furthermore, algorithmic latency, due to the overlap-add procedure from the ISTFT operation, equals to the frame length (e.g. 4 sub-frames for $25\%$ overlap, as used in this paper). In case of the original DCCRN, which also uses $25\%$ overlap, look ahead amounted to 6 sub-frames. Thus the baseline system requires relative $50\%$ algorithmic latency increase in reference to our proposed model.

\subsection{Further modifications to the network architecture and loss function}
\label{other_mods}
Further modification include an introduction of convolutional pathways, with filter kernel of $1 \times 1$, between the encoder and decoder through skip-connections, which is a similar idea to the DCCRN+ presented in \cite{dccrn+}. However, instead of performing concatenation along the feature dimension, we propose to sum the output from the convolution path with the output of the preceding deconvolutional layer, before feeding it to the next deconvolutional layer.

In the majority of experiments, we apply the typically used loss function~\cite{dccrn}, known as the scale-invariant signal-to-noise ratio (SI-SNR), which is defined as follows
\begin{equation}
\mathcal{L}_{\textrm{SI-SNR}} = 20\log_{10} \frac{{\lVert s \rVert}_{2}}{{\lVert \hat{s} - s \rVert}_{2}} \, ,
\end{equation}
where $s$ is scaled during training according to $s=(\hat{s}^T s) s/(s^T s)$ to normalize signal gain before loss computation. Note that the negative SI-SNR loss is minimized.

In the final experiments, we also modify the loss function as follows (based on the magnitude of the re-synthesized signal):
\begin{equation}
\label{eq:new_loss}
\mathcal{L}_{\textrm{SI-SNR+Mag}} = \gamma \mathcal{L}_{\textrm{SI-SNR}} + (1-\gamma) {\lVert \lvert \textrm{STFT}(\hat{s}) \rvert - \lvert \textrm{STFT}(s) \rvert \rVert}_{1} \, ,
\end{equation}
where hyperparameter $\gamma$ enables to weight the impact of both loss terms.
Note that the STFT used in the loss function differs from the one used to transform the input signal into the STFT domain. In particular, we apply the rectangular window, whilst all STFT parameters (such as frame length and hop-size) remain the same as in the processing path.

\section{Experimental Evaluation and Results}
\label{sec:eval}

\subsection{Datasets and training setup}

As a main dataset for the performed experiments, we used LibriMix (Libri2Mix) \cite{librimix}, precisely \emph{train-360} (50k utterances), \emph{dev} (3k utterances) and \emph{test} (3k utterances) splits in the \emph{min} mode. The training data consists of short (3\,s long) signals with 16\,kHz sampling frequency that undergo 512-point STFT with 32\,ms Hann window and 25\% (8\,ms) hop-size.
We base our implementation on the one available in the Asteroid toolkit \cite{asteroid}. We share most of the configuration and hyperparameters with Asteroid implementation\footnote{\url{https://github.com/asteroid-team/asteroid/tree/master/egs/librimix/DCCRNet}}, i.e Adam optimizer with weight decay set to 10e-5 and the maximum number of epochs set to 200 (using early stopping optimization with patience parameter equal 30). However, we increase the learning rate to 10e-2 and number of batches to 64, to fully utilize DDP training on 4 GPUs. In the preliminary experiments, we analyze training stability by performing four independent trainings for different seeds (that influenced parameters initialization and shuffling of batches), obtaining $0.024$ standard error for the SI-SDR measure \cite{SI-SDR} for model without any modifications.

In addition to the evaluation on the \emph{test} (3k utterances) split of LibriMix (Libri2Mix) \cite{librimix}, evaluation was also performed on synthetic test split (without reverb) from the Deep Noise Suppression (DNS) 2020 Challenge \cite{dns2020}, which consisted of 150 pairs of clean and noisy audio samples. Finally, as the values of the two losses in \eqref{eq:new_loss} are not in the same scale, we empirically set $\gamma = 0.995$.

\subsection{Performed experiments and evaluation metrics}

In experimental evaluation, we focus primarily on (i) the performance comparison between the mask-based and signal-based DCCRN, as well as (ii) comparison between two versions of the overlapped-frame prediction algorithm (with partial and full sub-frame summation) against the popular single-frame processing, and (iii) comparison between non-causal and causal architecture design. For the final setup, i.e. signal-based DCCRN with overlapped-frame prediction performing full-sub-frame summation and operating in a causal manner, we assess the model performance for the added convolutional pathways and the modified loss function given by \eqref{eq:new_loss}.

We used standard speech enhancement evaluation metrics such as the Scale-Invariant Signal-to-Distortion Ratio (SI-SDR) \cite{SI-SDR}, Perceptual Evaluation of Speech Quality (PESQ) \cite{PESQ}, and Short-Time Objective Intelligibility (STOI) \cite{STOI}. Table \ref{tab:final_res_libri_dns} presents the improvements (denoted as $\Delta$) of these measures between the processed (speech enhanced) and unprocessed (noisy) signals, for the evaluations performed on the Librimix and DNS datasets.

\begin{table*}[ht!]
\centering
\caption{Comparison of the original mask-based DCCRN with the proposed signal-based DCCRN for causal and non-causal single-frame and overlapped-frame prediction with full and partial summation. Results are presented for the LibriMix and DNS test datasets, with the reference values of \textDelta SI-SDR {[}dB{]}, \textDelta STOI, and \textDelta PESQ metrics (which are averaged over unprocessed signals for all datasets) equal to 3.4 dB, 0.796, 1.16 for Librimix, and 9.2 dB, 0.915, and 1.58 for DNS. Note that the algorithmic latency of causal vs non-causal cases is equal to 32 ms (4 sub-frames of 8 ms) vs 48 ms (6 sub-frames of 8 ms), respectively.}
\resizebox{1.85\columnwidth}{!}{%
\begin{tabular}{cclc<{\hspace{3pt}}ccc<{\hspace{3pt}}ccc}
\toprule
\multicolumn{1}{c}{} & \multicolumn{1}{c}{} & \multicolumn{1}{c}{} & \multicolumn{1}{c}{} & \multicolumn{3}{c}{\textit{Librimix}} & \multicolumn{3}{c}{\textit{DNS}} \\
\multicolumn{1}{c}{} & \multicolumn{1}{c}{\multirow{-2}{*}{Processing type}} & \multicolumn{1}{c}{\multirow{-2}{*}{Causal}} &  \multicolumn{1}{c}{\multirow{-2}{*}{\#Params (M)}}  & \textbf{\textDelta SI-SDR {[}dB{]}}    & \textbf{\textDelta STOI} & \textbf{\textDelta PESQ} & \textbf{\textDelta SI-SDR {[}dB{]}}    & \textbf{\textDelta STOI} & \textbf{\textDelta PESQ} \\ 
\midrule
\multicolumn{1}{c}{} & \multicolumn{1}{c}{} & no  & \textbf{3.7} & 10.0 & 0.121 & 1.04 & 7.2 & 0.046 & 1.02\\ 
\multicolumn{1}{c}{} & \multicolumn{1}{c}{\multirow{-2}{*}{single-frame pred.}}   & yes  & \textbf{2.8} & 9.7  & 0.116 & 0.95 & 7.0  & 0.045 & 0.96\\
\cmidrule{2-10}
\multicolumn{1}{c}{} & \multicolumn{1}{c}{} & no & 3.7& 10.1 & 0.123 & 1.07 & 7.3 & 0.048 & 1.04\\ 
\multicolumn{1}{c}{} & \multicolumn{1}{c}{\multirow{-2}{*}{\begin{tabular}[c]{@{}c@{}}overlapped-frame pred. \\ (partial sum.)\end{tabular}}} & yes  & 2.8 & 9.7 & 0.117 & 1.00 & 7.0 & 0.045 & 1.00\\
\cmidrule{2-10}
\multicolumn{1}{c}{} & \multicolumn{1}{c}{} & no & 3.7& 10.1 & 0.120 & 1.10 & 7.3 & 0.048 & 1.07\\ 
\multicolumn{1}{c}{\multirow{-7}{*}{\rotatebox[origin=c]{90}{\textit{mask-based}}}} & \multicolumn{1}{c}{\multirow{-2}{*}{\begin{tabular}[c]{@{}c@{}}overlapped-frame pred. \\ (full sum.)\end{tabular}}} & yes  & 2.8 & 9.7 & 0.114 & 1.03 & 7.0 & 0.045 & 1.03 \\ 
\midrule 
\multicolumn{1}{c}{} & \multicolumn{1}{c}{} & no & {\textbf{3.8}} & 10.2 & 0.123 & 1.06 & 7.3 & 0.047 & 0.99\\ 
\multicolumn{1}{c}{} & \multicolumn{1}{c}{\multirow{-2}{*}{single-frame pred.}} & yes  & {\textbf{2.9}} & 10.1 & 0.120 & 0.96 & 7.3 & 0.046 & 0.98\\
\cmidrule{2-10}
\multicolumn{1}{c}{} & \multicolumn{1}{c}{} & no & 3.8 & 10.6 & 0.128 & 1.10 & 7.6 & 0.049 & 1.06\\ 
\multicolumn{1}{c}{} & \multicolumn{1}{c}{\multirow{-2}{*}{\begin{tabular}[c]{@{}c@{}}overlapped-frame pred. \\ (partial sum.)\end{tabular}}} & yes  & 2.9 & 10.0 & 0.120 & 0.97 & 7.0 & 0.045 & 0.94\\
\cmidrule{2-10}
\multicolumn{1}{c}{} & \multicolumn{1}{c}{} & no & 3.8 & 10.5 & 0.127 & 1.10 & 7.4 & 0.048 & 1.04\\ 
\multicolumn{1}{c}{} & \multicolumn{1}{c}{} & yes  & 2.9 & 10.3 & 0.123 & 1.02 & 7.4 & 0.046 & 1.00\\ 
\multicolumn{1}{c}{} & \multicolumn{1}{c}{} & \multicolumn{1}{l}{{\xspace\xspace + CP}} & {\textbf{2.6}} & 10.2 & 0.122 & 1.00 & 7.3 & 0.047 & 0.99\\
\multicolumn{1}{c}{\multirow{-9}{*}{\rotatebox[origin=c]{90}{\textit{signal-based}}}} & \multicolumn{1}{c}{\multirow{-4}{*}{\begin{tabular}[c]{@{}c@{}}overlapped-frame pred. \\ (full sum.)\end{tabular}}} & \multicolumn{1}{l}{{\xspace\xspace\xspace + SI-SNR+Mag}} & {\textbf{2.6}} & 10.2 & 0.126 & 1.14 & 7.4 & 0.050 & 1.17\\
\bottomrule
\end{tabular}}
\label{tab:final_res_libri_dns}
\end{table*}

\subsection{Discussion of results}

The results presented in Table \ref{tab:final_res_libri_dns} for the Librimix dataset indicate that slight yet consistent improvement in SI-SDR and STOI is obtained by the proposed signal-based processing over the respective mask-based counterpart; improvement in terms of PESQ is also observed for the counterpart models.
For the DNS dataset, we can similarly observe improvement or at least the same gain for SI-SDR and STOI in the signal-based processing over the mask-based processing for the proposed causal as well as non-causal cases.

As expected, in general, switching from non-causal to causal processing, typically results in slight performance drop in terms of the evaluation metrics' values. On the other hand, it also significantly reduces the number of network parameters, by around $24\%$.

Comparing the processing type for the proposed causal signal-based processing, the presented overlapped-frame prediction with full sub-frame summation, consistently outperforms the standard single-frame prediction. Interestingly, partial summation notably outperforms the reference single-frame prediction for the non-causal processing. It should be noted, that comparing popular single-frame prediction in mask-based processing with the proposed signal-based processing with overlapped-frame prediction, in general performance gain is observed for the vast majority of cases, while at times minor hearable artifacts may also occur \cite{wang2022stft}. %

\begin{figure}[!t]
\centerline{\includegraphics[width=0.95\columnwidth]{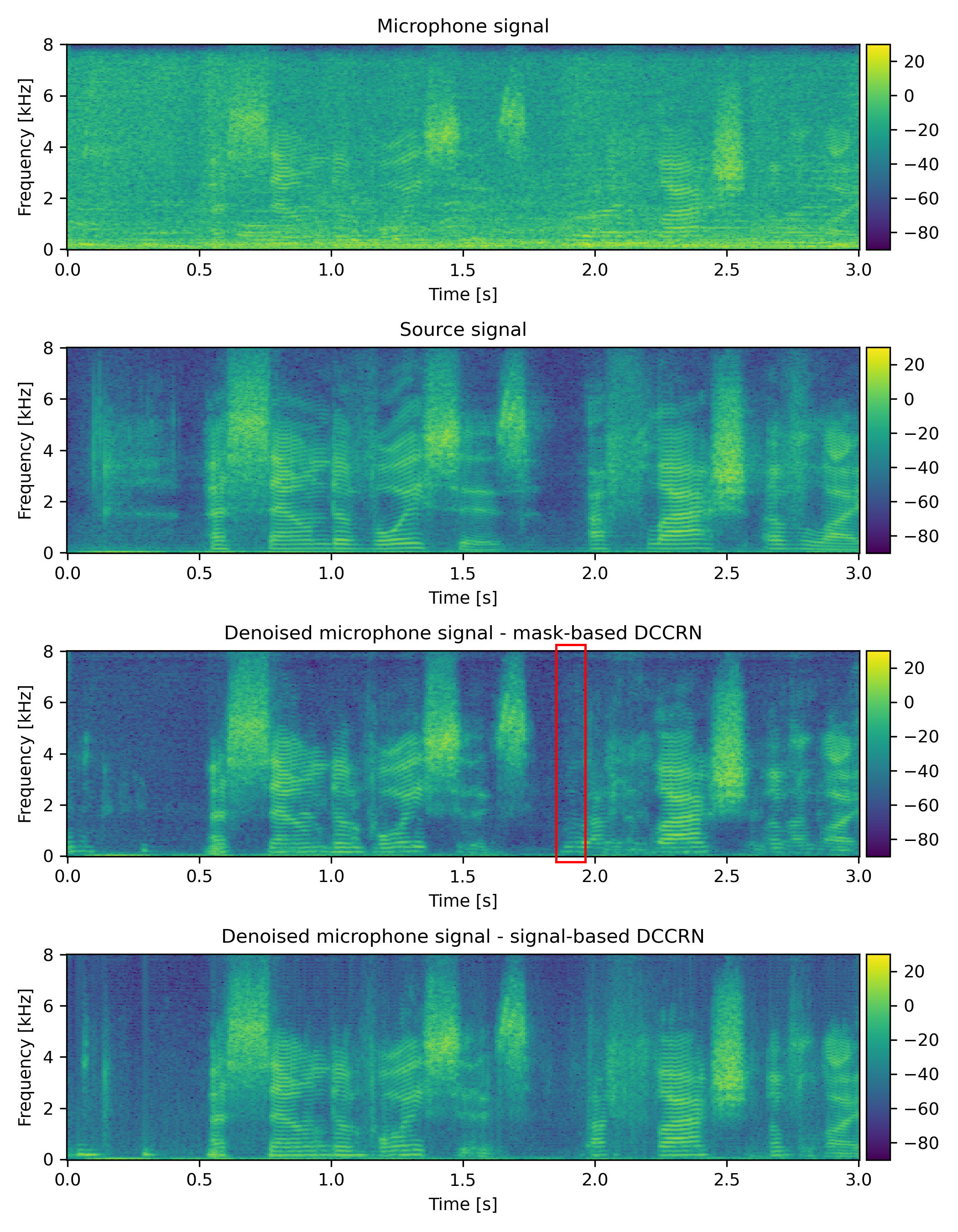}}
\caption{Example spectrograms of unprocessed microphone signal (top), source signal, source signal estimated by mask-based DCCRN, and signal-based causal DCCRN with convolutional pathways and applied overlapped-frame prediction using full sub-frame summation (bottom).}
\label{fig:specs}
\end{figure}

The proposed causal signal-based processing with full summation in overlapped-frame prediction, is next modified by adding convolutional pathways (CP). Such CP modification reduces the number of network parameters, while the evaluation metrics remain at very similar levels. 
Further modification of the loss function yields significant performance gain in STOI and PESQ, which is superior to the original mask-based DCCRN across all performance measures and both datasets. Note that this impressive results are obtained for the model with a reduced number of network parameters (from 3.7M to 2.6M).

Finally, Fig. \ref{fig:specs} compares example spectrograms of the original mask-based processing with single-frame prediction, and the proposed signal-based causal processing with overlapped-frame prediction using full sub-frame summation. As can be observed, spectrograms obtained with the proposed and state-of-the-art models are very similar. They are also similar to the spectrogram of the clean source signal, which indicates very good noise reduction. Nonetheless, some differences can be observed, e.g. in the time period just before 2 s, in which the existing mask-based approach estimates the signal which is not present in the original source signal. In contrast, spectrogram of the proposed model resembles more closely the spectrogram of the clean source in this time period.

\section{Conclusions}
\label{sec:conc}
This paper presents causal signal-based DCCRN for improved online single-channel speech enhancement. The proposed model performs direct complex filtering of the microphone signal with overlapped-frame prediction using full sub-frame summation. The results of experimental evaluation indicate that the proposed model achieves similar or even better speech enhancement than the original DCCRN with mask-based single-frame prediction, while it decreases the algorithmic latency and reduces the number of neural network parameters.

\section{Acknowledgements}
This research was funded in part by the National Science Centre, Poland DEC-2021/42/E/ST7/00452, by program `Excellence initiative – research university' for the AGH University of Science and Technology, and by the Foundation for Polish Science under grant number First TEAM/2017-3/23 (POIR.04.04.00-00-3FC4/17-00). For the purpose of Open Access, the author has applied a CC-BY public copyright licence to any Author Accepted Manuscript (AAM) version arising from this submission. The support of Poland's high-performance computing infrastructure PLGrid is acknowledged.

\bibliographystyle{IEEEtran}
\bibliography{mybib}

\end{document}